# The Nearby and Extremely Metal-Poor Galaxy

# CGCG 269-049[1]

Michael R. Corbin

U.S. Naval Observatory, Flagstaff Station, 10391 W. Naval Observatory Rd.,

Flagstaff, AZ, 86001-8521; *mcorbin@nofs.navy.mil*

Hwihyun Kim, Rolf A. Jansen, Rogier A. Windhorst

School of Earth and Space Exploration, Arizona State University, Tempe, AZ 85287

& Roberto Cid Fernandes

Departmento de Fisíca – CFM, Universidad Federal de Santa Catarina, C.P. 476, 88040-900

Florianópolis, SC, Brasil

---

ABSTRACT

We present *Hubble Space Telescope* (*HST*) and S*pitzer Space Telescope* images and photometry of the extremely metal-poor ($Z \cong 0.03 \, Z_\odot$) blue dwarf galaxy CGCG 269-049.  The *HST* images reveal a large population of red giant and asymptotic giant branch stars, ruling out the possibility that the galaxy has recently formed.  From the magnitude of the tip of the red giant branch we measure a distance to CGCG 269-049 of only $4.9 \pm 0.4$ Mpc.   The spectral energy distribution of the galaxy between ~3.6 µm – 70 µm is also best fitted by emission from predominantly ~10 Gyr old stars, with a component of thermal dust emission having a temperature of $52 \pm 10$ K.  The *HST* and *Spitzer* photometry indicate that more than 60% of CGCG 269-049's stellar mass consists of stars ~10 Gyr old, similar to other local blue dwarf galaxies.  Our *HST* Hα image shows no evidence of a supernova-driven outflow that could be removing metals from the galaxy, nor do we find evidence that such outflows occurred in the past.  Taken together with CGCG 269-049's large ratio of neutral hydrogen mass to stellar mass (~10), these results are consistent with recent simulations in which the metal deficiency of local dwarf galaxies results mainly from inefficient star formation, rather than youth or the escape of supernova ejecta.

*subject headings*:  galaxies: dwarf – galaxies: formation – galaxies: stellar content – galaxies individual: CGCG 269-049

## 1. INTRODUCTION

A small fraction of star-forming dwarf galaxies in the local universe have oxygen to hydrogen abundance ratios of $12 + \log(O/H) < 7.65$ in their emission-line gas, with an implied metallicity below 10% of the solar value (e.g., Kniazev et al. 2003; Izotov et al. 2006; Pustilnik & Martin 2007; Izotov & Thuan 2007). It is reasonable to hypothesize that such extremely metal-poor galaxies (XMPGs) are young objects, forming their first generation of stars, and considerable effort has recently been devoted to testing this hypothesis by studying the stellar populations of XMPGs including the prototype object, I Zw 18. While most XMPGs have been found to contain evolved stars (e.g., Aloisi et al. 2005; Corbin et al. 2006; see also the reviews by Kunth & Östlin 2000 and Kunth & Östlin 2007), the results for I Zw 18 itself have been controversial, with some studies based on deep *Hubble Space Telescope* imaging reporting no stars older than ~1 Gyr (Izotov & Thuan 2004; Östlin & Mouhcine 2005), while subsequent analyses of the Izotov & Thuan (2004) data and more recent *HST* observations indicate the presence of red giant stars in the galaxy (Momany et al. 2005; Aloisi et al. 2007). Both possibilities for the nature of XMPGs (evolved or non-evolved) have important implications for models of galaxy formation, which motivates the detailed study of more members of this class. In particular, the resolution of individual stars by *HST* imaging is the best way to establish the presence or absence of an evolved population. At shorter infrared wavelengths (< 10 μm), *Spitzer Space Telescope* photometry also constrains the age of their stellar populations, modulo the presence of Polyaromatic Hydrocarbon (PAH) emission. At longer infrared wavelengths, *Spitzer* photometry constrains the dust content of the galaxies, which both affects their optical photometry and carries implications for their star formation histories.

This paper presents deep *HST* and *Spitzer* images of the galaxy CGCG 269-049, identified as an XMPG by Kniazev et al. (2003). These authors measure a value of $12 + \log(\text{O/H}) = 7.43 \pm 0.06$ from the strong emission lines in its Sloan Digital Sky Survey (SDSS) spectrum, with an indicated metallicity of only 0.03 $Z_\odot$, assuming $\log(\text{O/H})_\odot = -3.08$ (Anders & Grevasse 1989). CGCG 269-049 is an isolated low-surface brightness dwarf galaxy with $g = 15.13$, $g - r \cong 0.03$, and a redshift of 0.00052 (Kniazev et al. 2003), placing it at a distance of $4.7 \pm 0.4$ Mpc after correction to the frame of the Cosmic Microwave Background (assuming $H_0 = 73$ km s$^{-1}$ Mpc$^{-1}$, $\Omega_M = 0.27$, $\Omega_\Lambda = 0.73$; Spergel et al. 2007). This distance presents an excellent opportunity to study CGCG 269-049 in detail. Most importantly, it allows the resolution of its stars with the *HST* Advanced Camera for Surveys / High Resolution Channel (ACS/HRC). The galaxy also has a major axis diameter of approximately 30″, allowing it to be resolved by the *Spitzer* Infrared Array Camera (IRAC) and Multiband Imaging Photometer (MIPS).

A color composite image of CGCG 269-049 from the SDSS is shown in Figure 1, which also shows the area of the galaxy imaged by the ACS/HRC. The galaxy shows star formation across most of its surface, and a population of old stars is not obviously present. Figure 1 shows that most of the star formation in CGCG 269-049 is concentrated in a star cluster offset to the southwest from its nominal center, which we hereafter refer to as the Central Cluster. H I observations of the galaxy reveal a semi-ordered velocity field and a symmetric H I envelope extending ≈ 2.3 Holmberg radii beyond the galaxy's optical disk (Begum et al. 2006). The symmetry of the optical and H I envelopes argues against the possibility that the galaxy's current star formation has been triggered by interaction with a neighbor, although it is only approximately 14.5 kpc from the dwarf galaxy UGC 7298, and is a member of the Canes Venatici I galaxy cloud (see Begum et al. 2006).

In the following sections we present our *Spitzer* and *HST* observations of CGCG 269-049, including a stellar color-magnitude diagram from the *HST* photometry that clearly shows the presence of asymptotic giant branch and red giant branch stars in the galaxy, with ages up to ~10 Gyr. We conclude with a discussion of these results in the context of recent models of the origin of the global galaxy mass / metallicity relation and dwarf galaxy formation. In the following calculations we assume a distance of 4.9 Mpc to the galaxy, measured from the magnitude of the tip of the red giant branch (TRGB) we detect in its color-magnitude diagram (§ 2.4). The corresponding scale is approximately 25 pc arcsec$^{-1}$.

## 2. OBSERVATIONS & RESULTS

### 2.1 *Spitzer Images and Photometry*

CGCG 269-049 was observed by the *Spitzer* IRAC and MIPS instruments on UT 12/24/2005 and 12/6/2005, respectively. The total integration times were 3701s (IRAC) and 6322s (MIPS). The galaxy was detected in all four IRAC bands (3.6 µm, 4.5 µm, 5.8 µm, and 8.0 µm) and in the MIPS 24 µm and 70 µm bands, but not the 160 µm band. Figure 2 shows the images of these galaxies in four of these bands. The 3.6 µm image is remarkably similar to the optical image (Figure 1). The images show a progressive decrease in the stellar flux of the galaxy in the regions away from the Central Cluster with increasing wavelength, and a gradual increase in the emission from the Central Cluster, which becomes strongest at 24 µm and 70 µm. This likely represents the decline of the stellar continuum emission beyond ~10 µm and the increase of thermal dust and PAH emission from the Central Cluster. Similar spatial correlation between emission above ~10 µm and star-forming regions detected at optical wavelengths is seen in *Spitzer* images of other nearby galaxies (e.g., Jackson et al. 2006; Barmby et al. 2006; Cannon et al. 2006; Hinz et al. 2006). No significant emission beyond the Central Cluster is detected in the MIPS bands.

Photometry was performed on these images with the IRAF "polyphot" task, using comparable polygonal apertures for all images to include the flux from the entire galaxy. The measured fluxes are given in Table 1. Figure 3 presents the resulting spectral energy distribution (SED). The SED is well-fitted by the combination of a Bruzual & Charlot (2003) stellar population synthesis model at the shorter wavelengths, and a Draine & Li (2007) thermal dust emission spectrum at longer wavelengths, including a significant contribution to the flux at 8.0 μm from PAH emission. The Bruzual & Charlot (2003) model that best fits the IRAC fluxes is an instantaneous burst with a Chabrier initial mass function, a metallicity of 0.02 $Z_\odot$, no internal extinction, and an age of 10 Gyr. We note, however, that the same model with an age of 1 Gyr also fits the data to within the estimated errors, so ages in the range of ~1 - 10 Gyr are possible, in addition to the newly-formed stars. This nonetheless provides the first evidence of old stars within the galaxy. The Draine & Li (2007) model that best fits the data has a dust composition based on that of the Large Magellanic Cloud, and a range of the parameter $U = 25 - 10^3$, where $U$ represents the intensity of the radiation field incident on the dust in units of the ambient radiation field in the solar neighborhood. The implied dust temperature from the fitted curve is $52 \pm 10$ K.

## 2.2  *Dust Content and Star Formation Rate*

We calculate the total mass of the warm dust in CGCG 269-049 from the fitted Draine & Li (2007) dust emission spectrum using their equation (31), and assuming a value of the ratio of the total dust mass to the total mass in hydrogen. We specifically use the mean value of this ratio, $2 \times 10^{-4}$, from the sample of 16 Blue Compact Dwarf Galaxies studied by Hirashita, Tajiri, & Kamaya (2002), which gives $M_{dust} \approx 2.3 \, M_\odot$. We note that for CGCG 269-049 this value of the dust-to-gas ratio is consistent with that estimated from the correlation between dust-to-gas ratio and the index $12 + \log(O/H)_{gas}$ found by Draine et al. (2007). The scatter in this correlation and in the Hirashita et al. (2002) values however make this mass estimate only suggestive; a larger value seems more appropriate to the strength of the 24 μm and 70 μm emission. We also find a low star formation rate (SFR) for the galaxy from its 8.0 μm luminosity using the calibration of Wu et al. (2005). We specifically find a SFR of approximately $10^{-5}$ $M_\odot$ yr$^{-1}$, noting the large uncertainty in this calibration (see Wu et al. 2005 and Calzetti et al. 2007). From the luminosity of the Hα emission-line in the galaxy's SDSS spectrum, we find a SFR of approximately $10^{-4}$ $M_\odot$ yr$^{-1}$ from the Kennicutt (1998) relation. The higher value of the SFR from the Hα luminosity is expected, given that the SDSS spectrum was obtained from a 3″ diameter fiber centered on the Central Cluster, whereas the 8.0 μm emission includes the outer regions of the galaxy, where the star formation is less intense.

As another check of the dust content of CGCG 269-049, we measure the Balmer decrement in its SDSS spectrum. We measure $f_\lambda$(Hα)/$f_\lambda$(Hβ) = 2.93 ± 0.05, with an associated $E(B\text{-}V)$ value of 0.056 mag, corresponding to $A_V \approx 0.2$ mag under the assumption of a Large Magellanic Cloud extinction curve (Gordon et al. 2003). We apply this value and associated $A_I$ and $E(V\text{-}I)$ values uniformly to correct our *HST* photometry for internal reddening, under the assumption that it does not vary strongly over the surface of the galaxy, given that we see no strong patches of extinction in Figure 1 and the *HST* images themselves.

## 2.3 *HST Images*

CGCG 269-049 was imaged by the *HST* Advanced Camera for Surveys on UT 11/2/2006 under GO program 10843. Broad-band images were taken with a single pointing of the High Resolution Channel, while an Hα image was taken in filter F658N with a single pointing of the Wide Field Channel. The filters and exposure times were F330W (2000s), F550M (6320s), F814W (2120s), and F658N (2192s). The broadband filters were chosen to cover a wide wavelength range, and to avoid strong emission lines including [O III] λλ 4959, 5007. The F658N filter includes Hα and the underlying stellar continuum, but this continuum should not contribute more than a few percent of the total flux based on the relative strength of the Hα line in the galaxy's spectrum (see Kniazev et al. 2003), at least near the Central Cluster. A separate off-band image was thus not taken. The exposure times of the broad-band images were chosen to detect individual red giants stars at the distance to the galaxy estimated from its redshift (§ 1). The raw data were processed with the standard ACS On-The-Fly-Recalibration Pipeline using the multidrizzling process. A correction for Galactic foreground extinction was applied to the image flux levels using the values of Schlegel et al. (1998).

A color composite of the F330W, F550M, and F814W images is shown in Figure 4. Individual stars of varying colors are clearly resolved. The Central Cluster is revealed to be more of a loose association dominated by the emission from ~20 OB stars. A faint background galaxy appears to be present approximately 6″ SE of the Central Cluster. This region was excluded from the stellar photometry, as was the area around the "coronagraphic finger" of the camera seen near the bottom of the figure, and the edges of the images.

A portion of the F658N image is shown in Figure 5. Hα emission is seen mainly around the Central Cluster, with what may be a single O star at the cluster center appearing to dominate the ionization. With the exception of its southwestern portion (possibly because of dust extinction), the emission is remarkably circular, indicative of a classical Strömgren sphere, with a diameter of approximately 175 pc. This contrasts with the filamentary appearance of ionized gas in "superwinds" from starburst galaxies (see the review by Veilleux, Cecil, & Bland-Hawthorn 2005). Although we lack data on the gas kinematics, the morphology and small size of the ionized volume do not indicate that a significant amount of gas is being driven from the Central Cluster. There is additionally no clear evidence of supernova remnants beyond the Central Cluster in this image.

### 2.4 *Photometry from HST Images*

Photometry of the resolved stars was performed on the F330W, F550M, and F814W images using the IRAF (version 2.12.2) implementation of DAOPHOT (Stetson 1987). An aperture size of 4 pixels (~ 0.1") was used. The stellar point-spread functions (PSFs) were modeled using the combination of an analytical function (Gaussian for the F330W and F550M filters, Moffat for the F814W filter) and an empirical composite PSF generated for each image. An iterative process of subtracting the detected stars and examining the residual image revealed that detection thresholds below $5\sigma$, where $\sigma$ is the standard deviation of the image background level, did not reliably detect additional stars. All magnitudes were placed on a Vega scale, using equation (4) of Sirianni et al. (2005) and updated zero-points from the ACS website (http://www.stsci.edu/hst/acs/analysis/zeropoints); they are 22.907, 24.392 and 24.861 for the F330W, F550M, and F814W images, respectively). These zero points assume an aperture of 5.5″, and our magnitudes were corrected accordingly, using the results of Sirianni et al. (2005).

Figure 6*a* presents a color-magnitude diagram (CMD) from the cross-matched lists of magnitudes in the F550M and F814W filters. We only plot magnitudes with formal 1$\sigma$ errors less than 0.2 mag. The CMD indicates the presence of a red giant branch, along with asymptotic giant branch stars and carbon stars. The stars near the 50% completeness limit shown on the plot are most likely to be dust-reddened post-AGB or carbon stars. These stars were visually verified to be genuine detections, and are distributed roughly uniformly over the portion of the galaxy imaged. We estimate a distance to the galaxy from the apparent F814W magnitude of the TRGB, 24.5 ± 0.1, and the assumption of M(*I*) = -4.05 for the TRGB of metal-poor systems (DaCosta & Armandroff 1990; see also Bellazzini, Ferraro, & Pancino 2001). Transforming the F814W magnitude to *I* (equation 12 and Table 23 of Sirianni et al. 2005) yields a distance of 4.9 ± 0.4 Mpc, in agreement with the value estimated from CGCG 269-049's redshift (§ 1). We overplot the theoretical isochrones calculated for ACS/HRC and these filters from the Padova database (http://pleiadi.pd.astro.it/isoc_photsy.02/isoc_acs_hrc/index.html; see also Girardi et al. 2002), assuming a distance of 4.9 Mpc and a metallicity of $Z$ = 0.0004. Good agreement is seen between the individual isochrones and the various branches of the CMD. Notably, the reddest RGB stars in the galaxy appear to be ~10 Gyr old. This CMD is similar to that obtained by Aloisi et al.

(2005) for the XMPG SBS 1415+437 using ACS/WFC and the F606W and F814W filters.

Figure 6*b* plots the F330W – F550M and F550M – F814W colors for the relatively small number of stars detected in the F330W filter. This comparison provides a check on the accuracy of the isochrone fitting to the CMD by including the F330W magnitudes. Specifically, the blue horizontal branch loops in the isochrones seen in Fig. 6*a* produce a nearly linear feature in this color space that fits the data very well. If systematic errors were present in our photometry or in the isochrone fitting they would likely be evident on one or both of Figures 6*a* and 6*b*, but the correspondence of the data and isochrones indicates otherwise.

To assess the completeness of the photometry for each filter, we follow a procedure similar to Caldwell (2006) for ACS observations of dwarf galaxies in the Virgo cluster. We specifically insert 100 artificial stars with empirical composite PSFs at random positions into the actual images for all three filters. The magnitude of the inserted stars was varied from 21 to 29 in steps of 0.25 mag, and a total of 544 trials of the procedure were run for each filter. The photometry of the artificial stars was performed with the same procedure applied to the actual stars. The results are shown in Figure 7, where it can be seen that the 50% (90%) completeness levels are reached at magnitudes of approximately 25.3, 27.2, 26.7 (24.8, 26.6, 26.0) for the F330W, F550M, and F814W filters, respectively. The F330W image is not deep enough to recover stars to the faint limits reached in the F550M and F814W images. Of the three images, F550M is effectively the deepest, with a completeness limit approaching 27 mag, while the F814W image reaches ~0.5 mag brighter. The number of stars detected to within an error of 0.2 mag are 309 (F330W), 3128 (F550M), and 4339

(F814W).

### 2.5 *Mass Fraction of Old Stars*

It is of interest to estimate the fraction of stellar mass represented by the stars in CGCG 269-049 older than ~1 Gyr to assess its star formation history.  Corbin et al. (2006) find that the ~10 Gyr old populations in nearby "ultracompact" blue dwarf galaxies, including two that qualify as XMPGs, comprise ~90% of their stellar mass, with their light being dominated by the early-type stars formed in their starbursts.  These estimates were however based on a decomposition of the objects' SDSS optical spectra into their constituent stellar populations, obtained through apertures containing nearly all of their emission.  The SDSS spectrum of CGCG 269-049 is centered on the Central Cluster, and is thus not representative of the galaxy.  We must consequently use the *HST* and *Spitzer* photometry to make this estimate.  We note that the relative lack of depth of the *HST* images and their incomplete coverage of the galaxy (Figure 1), as well as the uncertainties inherent in the CMD,  make these estimates only indicative.

For the *HST* photometry, we begin by noting that stars at the main-sequence turnoff in a ~10 Gyr population will have a mass ~1 $M_\odot$.  Their combined post main-sequence stages last ~500 Myr, most of which is spent on the RGB.  Hence the masses of the ~10 Gyr stars we detect should be in the range ~1 – 1.02 $M_\odot$.  Stars on the main-sequence turnoff in a ~1 Gyr population, by contrast, have masses ~2.5 $M_\odot$ and linger less than ~50 Myr on the RGB, making it probable that the RGB stars we detect have ages closer to 10 Gyr.  We count 1432 stars on the red branch of our CMD (having F550M – F814W > 0.4 mag and F814W < 27.0 mag).  Correcting for incompleteness at 26 < F814W < 27 mag and for RGB stars with 27 < F814W < 31.3 mag (the magnitude of the main-sequence turn-off), the total number of evolved stars must be ~3 times larger.  To correct for the number of evolved stars outside the ACS/HRC field of view, we use a ground-based $R$-band image CGCG 269-049 obtained with the 1.8m Vatican Advanced Technology Telescope on 2006 May 5.  This image has an integration time of 600s and a plate scale of 0.375″ pix$^{-1}$.  We find from an analysis of the radial brightness profile of this image that approximately 90% of the galaxy's total light is contained within the ACS/HRC field of view, which is consistent with inspection of Figure 1.  With these corrections we estimate the total number of RGB stars to be approximately 4,700.  Direct integration of the Miller & Scalo (1979) and Scalo (1986) initial mass functions implies that the total zero age main-sequence mass of the ~10 Gyr old population is 200 – 300 times larger than the total mass of stars in the ~1 – 1.02 $M_\odot$ range.  With our estimated number of RGB stars, this yields a total mass of approximately $1.00 – 1.50 \times 10^6$ $M_\odot$, where the range represents the difference between the Scalo (1986) and Miller & Scalo (1979) initial mass functions, both with lower and upper mass limits of 0.1 and 125 $M_\odot$.

We similarly estimate the masses of stars younger than ~1 Gyr from the *HST* photometry by counting the number of stars on the blue branch of the CMD (having F550M – F814W < 0.4 mag and F814W < 27); we find 1104.  Their distribution in the CMD and color-color diagram suggests that they belong to populations with ages of 250 Myr and younger. For corresponding main-sequence turnoff masses between 4.0 and 20 $M_\odot$ we estimate a total zero-age main sequence mass of $0.13 - 1.03 \times 10^6$ $M_\odot$. Changing the upper mass limit to 100 $M_\odot$ decreases this mass by less than 10%. The total mass in stars of CGCG 269-049 is thus ~$1.09 - 2.53 \times 10^6$ $M_\odot$, of which ~60% - 90% is contained in stars older than 1 Gyr. This fraction is consistent with the findings of Corbin et al. (2006) and with Aloisi et al. (2005).

We estimate the mass of the ~10 Gyr old stellar population from the *Spitzer* photometry by calculating the *V*-band luminosity of the best-fitting Bruzual & Charlot (2003) model to the IRAC fluxes (Figure 3). This method importantly avoids the contribution to the actual *V*-band flux from OB stars and nebular emission lines. The associated $M/L_V$ ratio for this model is approximately 2 (see Figure 1 of Bruzual & Charlot 2003), which yields a total mass for the ~10 Gyr old population of $2 \times 10^7$ $M_\odot$. This value is significantly higher than that found from the *HST* photometry, but is likely more accurate because of the more direct nature of the estimate. It also strengthens the conclusion that the majority of stars in CGCG 269-049 are ~10 Gyr old.

### 2.6 *Evidence of Past Outflows?*

One likely contribution to galaxy metal deficiency is the escape of supernova ejecta into the intergalactic medium (see e.g., Martin, Kobulnicky, & Heckman 2002). This in part motivated obtaining an Hα image of CGCG 269-049 in addition to broad-band images. While the Hα image (Figure 4) does not indicate a current outflow, such outflows may have occurred in the past, during more vigorous episodes of star formation. Evidence of such an outflow from the blue compact dwarf galaxy IC 691 has been reported by Keeney et al. (2006), who find Lyα and C IV λ1548,1551 absorption lines near the redshift of the galaxy in the ultraviolet spectrum of background QSO with a projected separation of 33 kpc from the galaxy. Motivated by this result, we searched the NASA Extragalactic Database for QSOs near CGCG 269-049 that could be used as probes of past outflows. We find one object, SDSS  J121507.48+522055.7 ($z = 1.21$) located approximately 6′ from CGCG 269-049 (a projected separation of 8.6 kpc) and at a position angle of approximately 230°, which is roughly aligned with its minor axis. However, we find no evidence of absorption lines (e.g., Ca II H & K) in the SDSS optical spectrum of this QSO that could arise from a past outflow from CGCG 269-049. Thus, either such an outflow did not occur, or else its remnants along the line of site to the QSO have a velocity dispersion and/or column density below the sensitivity of the SDSS spectrum (which has a resolution of approximately 140 km s$^{-1}$ and is moderately noisy). The lack of evidence for supernova remnants in the disk of CGCG 269-049 (Figure 5) argues in favor of the former possibility.

## 3. DISCUSSION

The clear presence of asymptotic giant branch and red giant branch stars in CGCG 269-049 revealed by these data rules out the possibility that it is a newly-formed galaxy. In addition to the studies discussed in § 1, these results argue strongly against this possibility for XMPGs in general. While the mean stellar ages among star-forming dwarf galaxies will be skewed to lower values than their quiescent counterparts, studies of extremely metal-poor dwarf galaxies in the Local Group without active star formation have shown them to consist mainly of evolved stars (see, e.g., Schulte-Ladbeck et al. 2002; Momany et al. 2005; Walsh, Jerjen, & Willman 2007). Among star-forming XMPGs such as CGCG 269-049, it appears that their current starbursts dominate their light and create a "blue façade" over stellar populations that are predominantly ~10 Gyr old (e.g., compare Fig. 1 and Fig. 4; see also Corbin et al. 2006). For more distant XMPGs whose stars cannot be resolved by *HST*, e.g., SBS 0335-052W (see Izotov, Thuan, & Guseva 2005), the best way to test for the presence of evolved stars is by the decomposition of their spectra into constituent stellar populations, using codes such as STARLIGHT (see Asari et al. 2007 and references therein). We note that in contrast to SBS 0335-052 (see Houck et al. 2004), the mid-infrared SED of CGCG 269-049 (Fig. 3) shows evidence for significant PAH emission. This agrees with evidence that the strength of PAH features among star-forming dwarf galaxies depends more strongly on star-formation rate than metallicity (Rosenberg et al. 2004; Jackson et al. 2006), noting that SBS 0335-052 has six "super" star clusters and an inferred star formation rate twice that of what we estimate for CGCG 269-049 (see Thuan, Izotov, & Lipovetsky 1997; § 2.2).

Brooks et al. (2007) have recently modeled the galaxy mass / metallicity relation (see Tremonti et al. 2004; Lee et al. 2006) using smoothed particle hydrodynamics and *N*-body simulations of galaxy formation including supernova feedback and associated metal enrichment. Their simulations extend down to metallicities as low as that measured for CGCG 269-049. They find that inefficient star formation among low-mass galaxies, rather than galaxy age or metal loss from supernova ejecta, is the main cause of the relation (but see Martin et al. 2002; Keeney et al. 2006; Davé et al. 2006; and Calura et al. 2007 for observational evidence and theoretical arguments for metal loss from supernova ejecta). The results for CGCG 269-049 are consistent with these simulations, because we find both an old population of stars and no strong evidence for current or past outflows. In addition, CGCG 269-049 has a high ratio of H I to stellar mass, indicative of an inefficient conversion of gas to stars. Specifically, the Begum et al. (2006) H I flux of CGCG 269-049 corresponds to an H I mass of approximately $3 \times 10^7 \, M_\odot$, which when compared to our estimate of the total stellar mass in the galaxy (§ 2.5) yields a ratio of H I to stellar mass ~10. This value is comparable to that found for the low surface-brightness dwarf galaxy ESO 215-G?009 by Warren, Jerjen, & Koribalski (2004), assuming a galaxy mass-to-light ratio ~2-4. The Brooks et al. (2007) simulations are also consistent with the finding of Kannappan (2004) that galaxy stellar mass and the ratio of H I mass to stellar mass are inversely correlated over several decades in both quantities. Specifically, this correlation would lead to low O/H gas abundance ratios in dwarf galaxies because of their higher relative H I content. Using the H I mass value from Begum et al. (2006) and our stellar mass estimate, CGCG 269-049 falls roughly on the Kannappan (2004) correlation, as well as the galaxy mass / metallicity correlation of Tremonti et al. (2004) as calibrated by Brooks et al. (2007).

While the present data do not offer a simple picture of CGCG 269-049's star-formation history, the result that the majority of its stars and those of similar galaxies (see Corbin et al. 2006) are ~10 Gyr old is broadly consistent with models of the formation of isolated dwarf galaxies in which the photo-evaporation of baryons from their dark matter halos at the epoch of reionization strongly lowers their star formation rates, leading to galaxies at the present epoch comprised mainly of old stars with some residual star formation (Hoeft et al. 2006; Wyithe & Loeb 2006).

We thank Bruce Draine for giving us access to his dust emission spectra prior to publication, and Seth Cohen for his assistance with the *HST* data. We also thank an anonymous referee for suggestions that improved the paper, and Bev Smith, Helmut Jerjen, Alyson Brooks, and Bill Vacca for helpful discussions. This work was supported by NASA under *Hubble Space Telescope* grant GO-10843.04-A and *Spitzer Space Telescope* grant RSA-1281759 to the United States Naval Observatory and Arizona State University. This work is based in part on observations with the Vatican Advanced Technology Telescope, and has used the NASA Extragalactic Database (NED), which is operated by the Jet Propulsion Laboratory of the California Institute of Technology, under contract with NASA.

TABLE 1

CGCG 269-049 SPITZER SPACE TELESCOPE IRAC AND MIPS FLUXES

| Wavelength, μm | Detector | $f_\nu$ (mJy) |
|---|---|---|
| 3.6 | IRAC | 1.23 ± 0.05 |
| 4.5 | IRAC | 0.82 ± 0.05 |
| 5.8 | IRAC | 0.54 ± 0.05 |
| 8.0 | IRAC | 0.44 ± 0.05 |
| 24 | MIPS | 0.60 ± 0.03 |
| 70 | MIPS | 6.08 ± 0.03 |

FIGURE CAPTIONS

Figure 1. - Optical image of CGCG 269-049 from the Sloan Digital Sky Survey (SDSS), representing a composite of images in the SDSS *g*, *r*, and *i* bands. The red rectangle shows the area covered by the *HST* Advanced Camera for Surveys / High Resolution Channel images (Figure 4 and § 2.3).

Figure 2. - *Spitzer Space Telescope* IRAC 3.6 µm and 8.0 µm band images, and MIPS 24 µm and 70 µm images of CGCG 269-049. The galaxy is also detected in the IRAC 4.5 µm and 5.8 µm bands, with an appearance intermediate between that in the 3.6 µm and 8.0 µm bands. Note the similarity of the 3.6 µm image to the optical image (Figure 1).

Figure 3. - Infrared spectral energy distribution of CGCG 269-049 from the *Spitzer* IRAC and MIPS fluxes, shown with open circles and solid squares, respectively. The solid line is a Bruzual & Charlot (2003) population synthesis model for an instantaneous burst evolved to an age of 10 Gyr, a Chabrier initial mass function, and a metallicity of 0.02 $Z_{\odot}$. The dashed line is a Draine & Li (2007) thermal dust model for a Large Magellanic Cloud dust composition and a range in the parameter $U = 25 - 10^{3}$, where $U$ represents the intensity of the radiation field incident on the dust in units of the ambient radiation field in the solar neighborhood. The dotted line is the sum of the two models.

Figure 4. - Color composite of the *HST* Advanced Camera for Surveys / High Resolution Channel F330W, F550M, and F814W images of CGCG 269-049. The feature at the bottom of the image is the "coronagraphic finger" of the ACS/HRC. The J2000 coordinates of the image center are RA = 12h 15m 46.89s, DEC = +52° 23′ 13.9″.

Figure 5. - A portion of the Hα image of CGCG 269-049 obtained with the F658N filter of the Advanced Camera for Surveys / Wide Field Channel. The image is centered on the Central Cluster (see Figs. 1 and 4).

Figure 6. - (*a*) Color-magnitude diagram of CGCG 269-049 from the photometry of the F550M and F814W images. Magnitudes are on a Vega scale. Padua isochrones for these filters and for a metallicity of $Z = 0.0004$ are plotted (see text for details). The dashed line represents the 50% completeness level. (*b*) Color-color diagram for the stars detected in the F330W, F550M, and F814W filters, with the same isochrones as in 6(*a*). The roughly linear portions of the isochrones on this plot correspond to the blue horizonal branch loops on 6*a*.

Figure 7. - Completeness of stellar photometry as a function of magnitude for each of the three filters used, as determined by inserting artificial stars into the images (see text for details). The error bars

represent the variance between different simulations, where stars were inserted at different positions in the images.

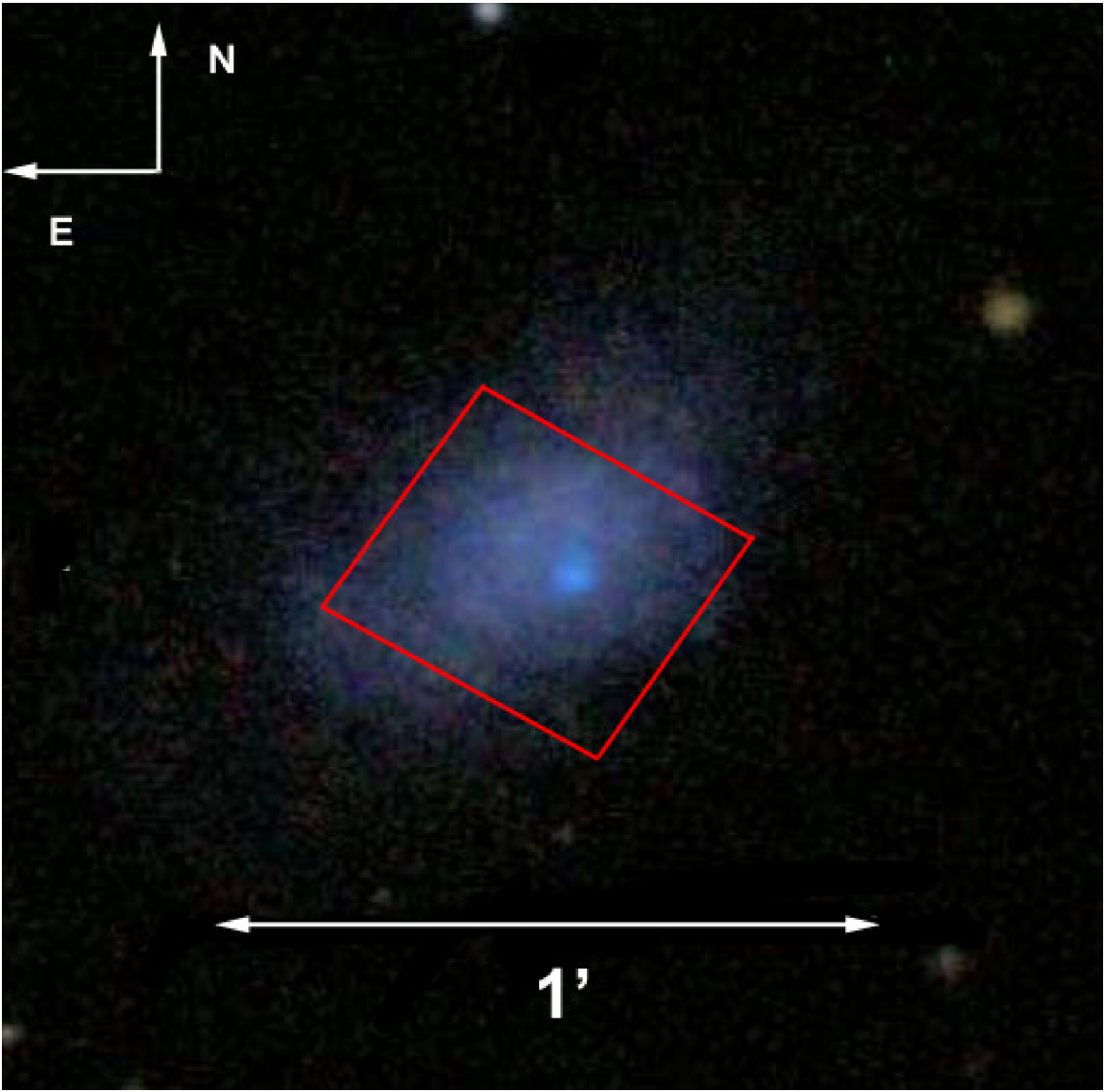

Figure 1.

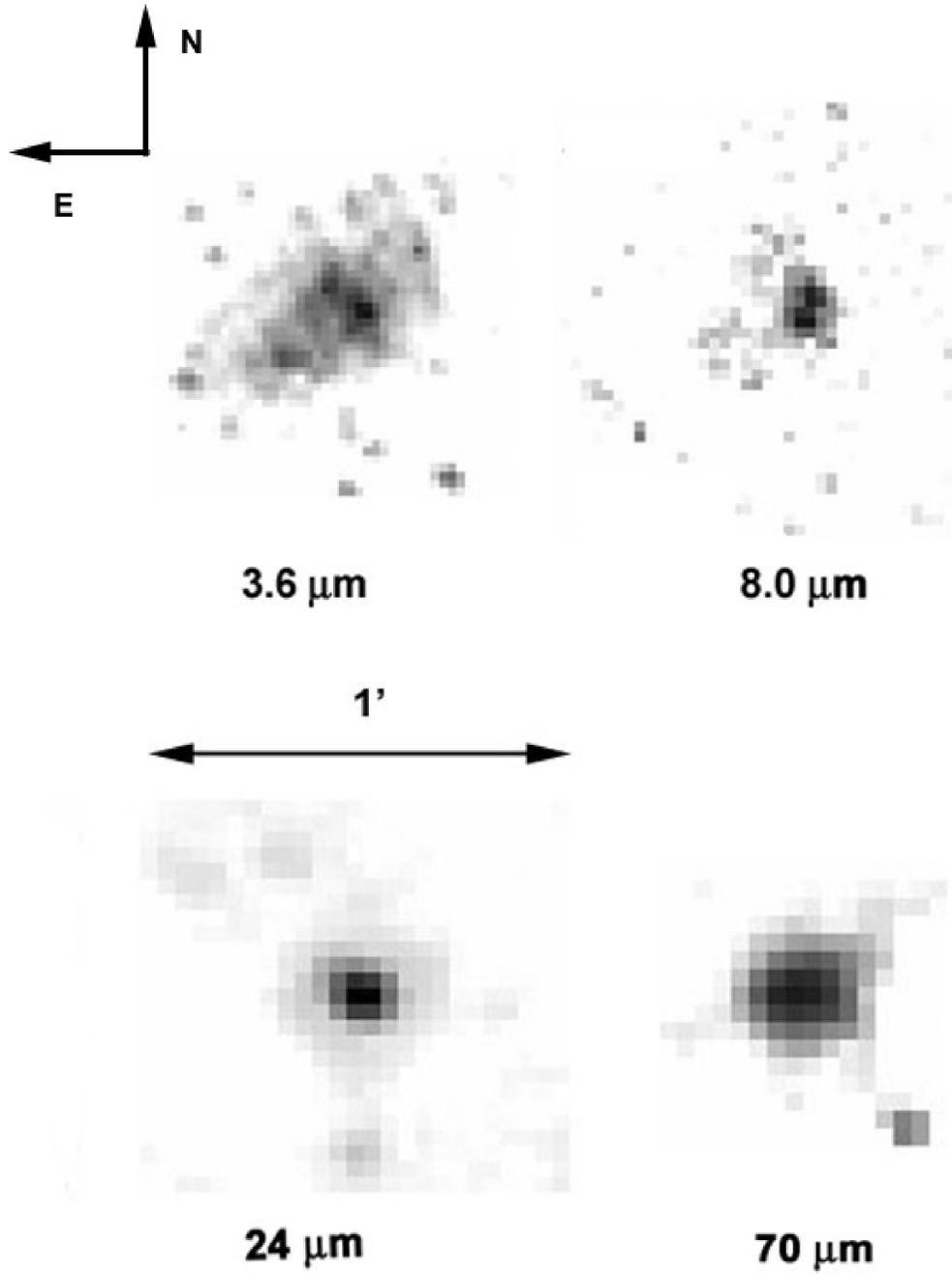

Figure 2.

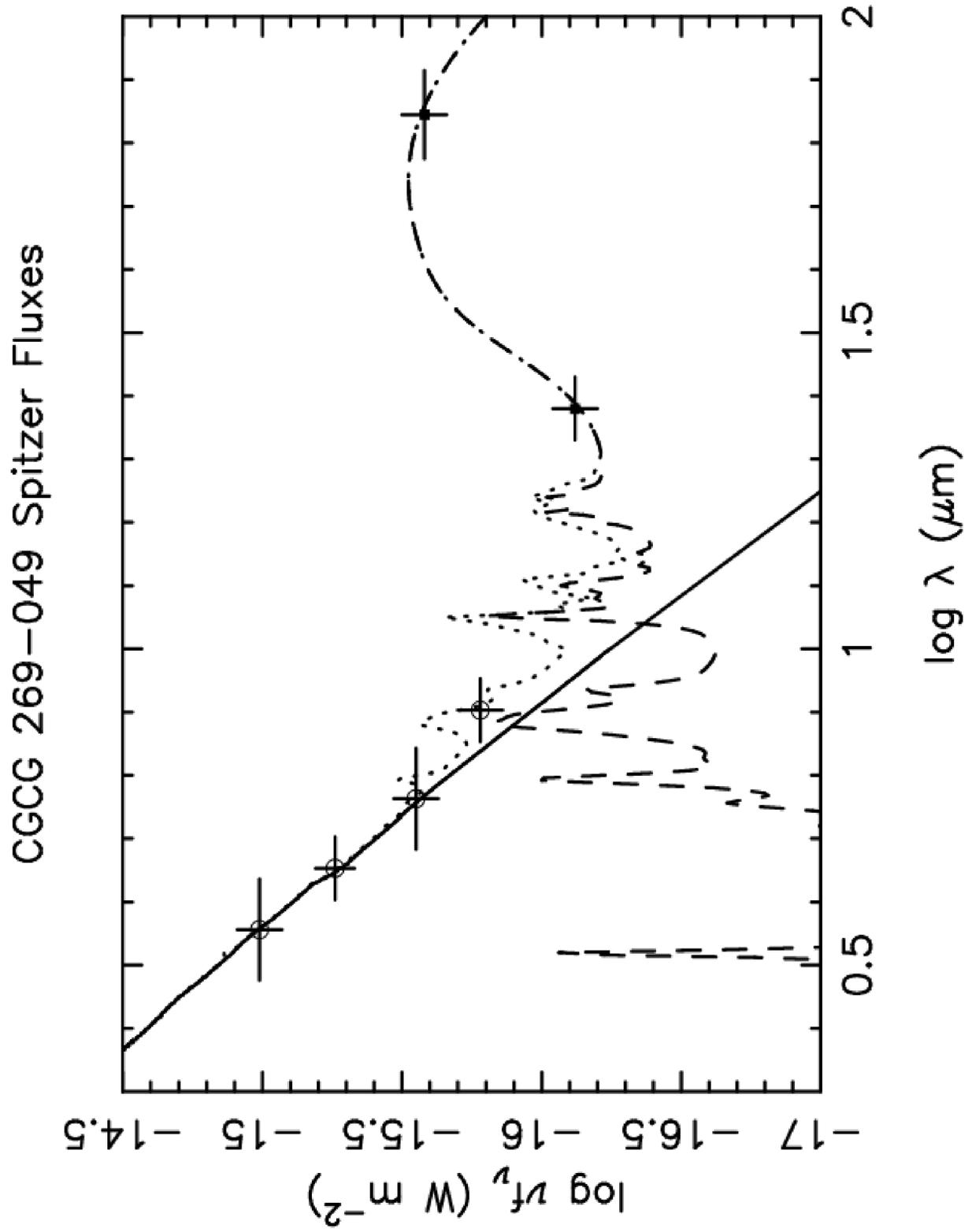

Figure 3.

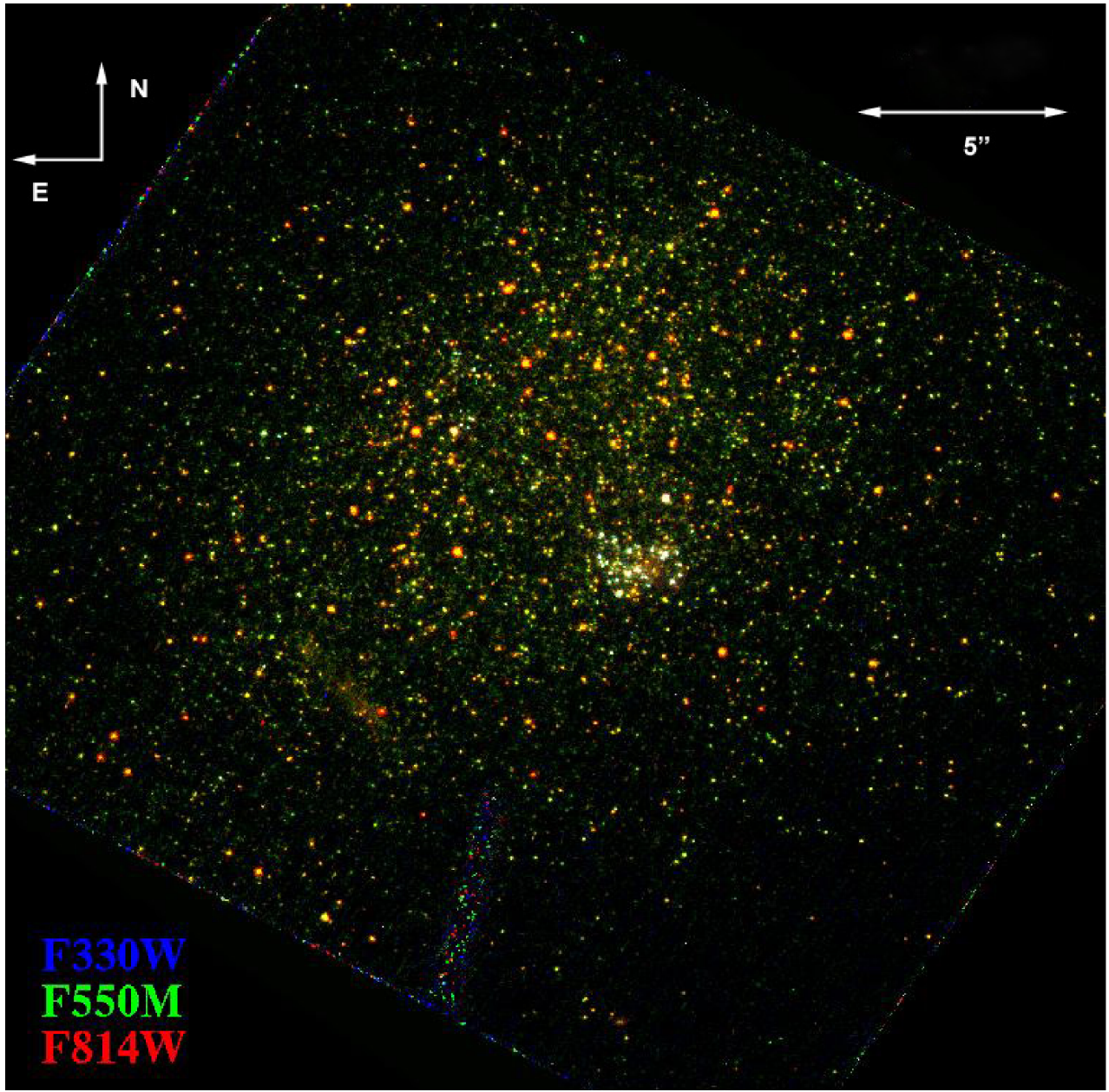

Figure 4.

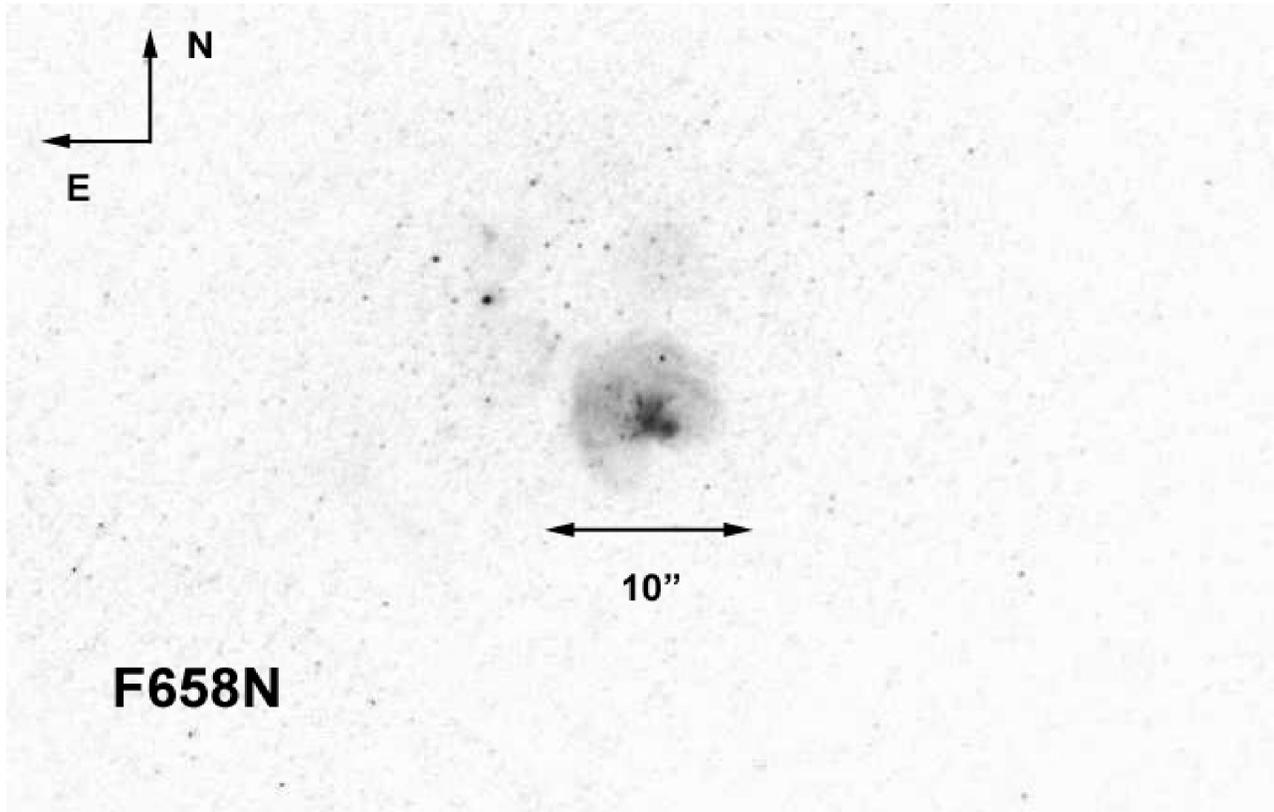

Figure 5.

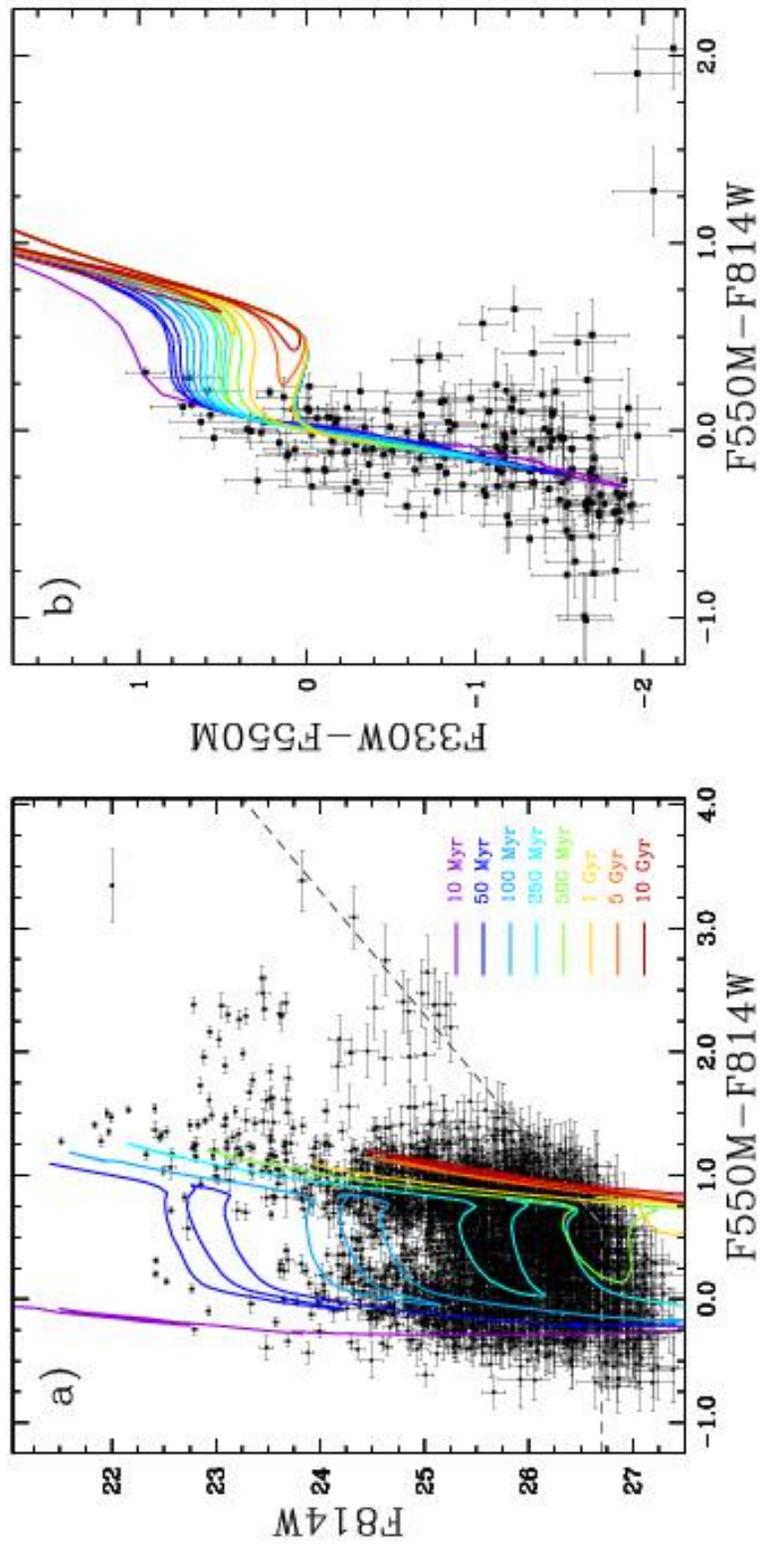

Figure 6.

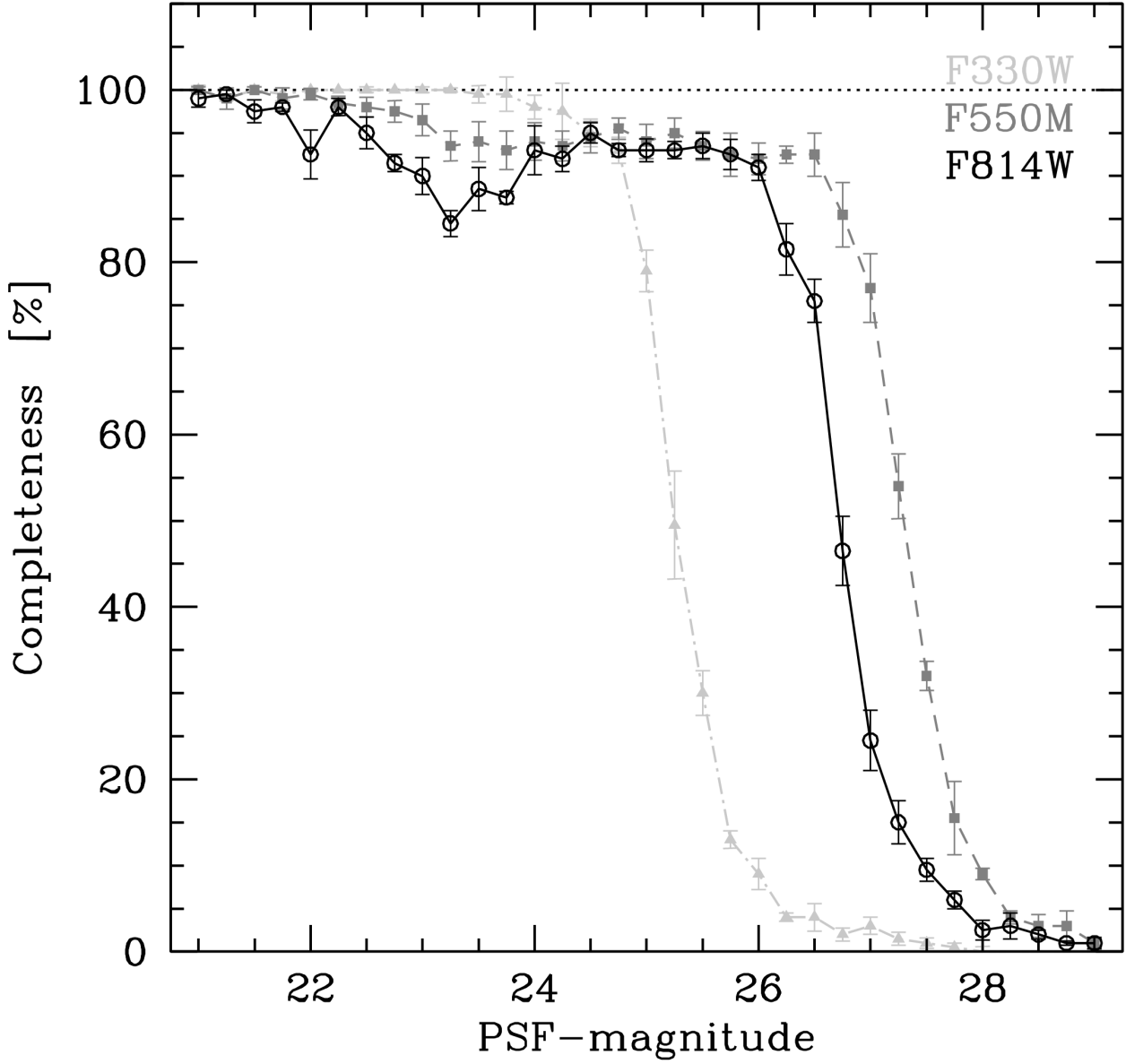

Figure 7.